\documentclass[runningheads]{llncs}
\usepackage[T1]{fontenc}
\usepackage{graphicx}
\usepackage{amsmath}
\usepackage{amsfonts}
\usepackage{cleveref}
\newcommand{\opp}[2]{\ensuremath{\operatorname{#1}\left( #2 \right)}}
\newcommand{\op}[1]{\ensuremath{\operatorname{#1}}}
\newcommand{\ie}{\textit{i.e.}}
\newcommand\at[1]{
{#1}{\raisebox{-0.5\height}{\(\left. \vphantom{#1} \right |\)}\vphantom{#1}}
}

\renewcommand{\orcidID}[1]{}

\begin{document}
\title{Defects in unidimensional structures}
%
%
\author{Mewen Crespo\inst{1,3}\orcidID{0000-0003-3834-9462} \and
Guy Casale\inst{1,4}\orcidID{0000-0002-3053-3102} \and
Lo\"ic Le Marrec\inst{1,5}\orcidID{0000-0002-1484-4971} \and
Patrizio Neff\inst{2,6}\orcidID{0000-0002-1615-8879}}
\authorrunning{M. Crespo et al.}
%
\institute{Univ Rennes, CNRS, IRMAR - UMR 6628, F-35000\\
\and Head of Chair for Nonlinear Analysis and Modelling, Faculty of Mathematics, University of Duisburg-Essen, Thea-Leymann-Straße 9, D-45127 Essen, Germany\\
\and \email{mewen.crespo@univ-rennes.fr}
\and \email{guy.casale@univ-rennes.fr}
\and \email{loic.lemarrec@univ-rennes.fr}\and
\email{patrizio.neff@uni-due.de}}
\maketitle              
\begin{abstract}
    In a previous work of the first authors, a non-holonomic model, generalising the micromorphic models and allowing for curvature (disclinations) to arise from the kinematic values, was presented. In the present paper, a generalisation of the classical models of Euler-Bernoulli and Timoshenko bending beams based on the mentioned work is proposed. The former is still composed of only one unidimensional scalar field, while the later introduces a third unidimensional scalar field, correcting the second order terms. The generalised Euler-Bernoulli beam is then shown to exhibit curvature (\ie{} disclinations) linked to a third order derivative of the displacement, but no torsion (dislocations). Parallelly, the generalised Timoshenko beam is shown to exhibit both curvature and torsion, where the former is linked to the non-holonomy introduced in the generalisation. Lastly, using variational calculus, asymptotic values for the value taken by the curvature in static equilibrium are obtained when the second order contribution becomes negligible; along with an equation for the torsion in the generalised Timoshenko beam.

\end{abstract}

\section{Introduction}

Continuum mechanics seeks to describe the behaviour of materials by treating them as continuous media governed by smooth field equations. Classical linear elasticity, based on small deformations and displacements, models the material body as a differentiable manifold embedded in Euclidean space, with the internal energy depending solely on the symmetric part of the displacement gradient. While highly effective for many engineering applications, such models often fall short when internal structure or scale effects become significant.

To capture these phenomena, generalized continuum theories have introduced additional kinematic variables. Second gradient models incorporate the Hessian of the displacement, accounting for microstructural interactions, but are limited to systems where the microstructure is a smooth function of the displacement. However, many materials exhibit non-integrable microstructures, characterised by the presence of defects, namely dislocations and disclinations~\cite{VolterraTheoryDislocations1907}. This motivates a broader framework where the microstructure is modelled independently.

The \emph{micromorphic model}, introduced by Mindlin, Eringen, et al.~\cite{MindlinMicrostructureLinearElasticity1964,EringenSuhubiNonlinearTheorySimple1964,ToupinElasticMaterialsCouplestresses1962}, extends classical elasticity by enriching the kinematics with an independent microstructural field. In this framework, in the linear setting, each material point carries not only a displacement \( u \), but also an internal deformation described by a second field \( P \), typically interpreted as a second-order tensor capturing micro-deformations~\cite{NeffEtAlUnifyingPerspectiveRelaxed2014,NeffForestGeometricallyExactMicromorphic2007}. The energy depends on both \( P \) and its gradient \( \nabla P \), allowing the model to incorporate characteristic length scales and capt    ure size-dependent effects. When \( P \) is allowed to vary freely (\ie{} without the constraint \( P = \nabla u \)), the model is not \emph{holonomic}, enabling it to account for defects such as dislocations, represented by \(\op{curl} P \neq 0 \).

However, if dislocations arise from a violation of Schwartz equalities for the first-order \(\op{curl}{\nabla u} = 0\), disclinations arise from a violation of Schwartz equalities for the second order: \(\opp{curl}{\nabla^2 u} = 0\) (see \cite{CartanGeneralisationNotionCourbure1922}). While the former can happen in micromorphic models, the later yields \(\op{curl}{\nabla P} = 0\), which is always true. In recent works \cite{CrespoEtAlTwoScaleGeometricModelling2024,CrespoEtAlContinuumMechanicsDefective2023}, a general variational model was proposed to describe such non-holonomic microstructured media using three fields: the macroscopic displacement \(u\), the microstructural distortion tensor \(P\), and a connection-like field \(N\) encoding relative microstructural distortion (\ie{} relaxing \(\nabla P\)). Furthermore, the measures of deformation, on which the energy must depend, have been computed in \cite{CrespoEtAlTwoScaleGeometricModelling2024}. In the linear settings, those reduces to: \(\nabla u - P\), \(\op{sym}{P}\) and \(N\).

In this paper, we explore a simplified, linearized version of this model suitable for analytical and numerical study and apply it to two classical beam models: Euler-Bernoulli and Timoshenko beams. Using variational techniques, we then derive the governing equations for the defect fields. As this is an exploratory study, we place ourselves in a bi-dimensional setting for simplicity.

\section{The bidimensional linear second-order non-holonomic model}

\subsection{Relation to non-linear formulations}

In two dimensions, when the macroscopic and microscopic scales are considered as independent dimensions, the kinematic spaces of \cite{CrespoEtAlTwoScaleGeometricModelling2024} can be locally identified with \(\mathbb{R}^2 \times \mathbb{R}^2\). Let us denote a typical point of the body by \(
\begin{bmatrix} \overline{X} & Y \end{bmatrix}^\mathrm{T} \in \mathbb{R}^2 \times \mathbb{R}^2
\).
In this context, the three kinematic variables \(-\) namely the macroscopic displacement gradient \(\nabla_{\overline{X}} u : \mathbb{R}^2 \to \mathbb{R}^2\), the microscopic deformation \(\at{P}_{\overline{X}} : \mathbb{R}^2 \to \mathbb{R}^2\), and the mixed-scale field \(\at{N}_{\begin{bmatrix} \overline{X} \\ Y \end{bmatrix}} : \mathbb{R}^2 \to \mathbb{R}^2\) \(-\) can be combined into a single linear operator \(\delta \mathbf{F} := \begin{bmatrix} \nabla u & \mathbf{0} \\ N & P \end{bmatrix}\). This representation is to be compared with the non-linear first-order placement tensor given in~\cite{CrespoEtAlTwoScaleGeometricModelling2024}: \(\mathbf{F} = \begin{bmatrix} \mathbf{F}_{\mathrm{h}}^{\mathrm{h}} & \mathbf{0} \\ \mathbf{F}_{\mathrm{h}}^{\mathrm{v}} & \mathbf{F}_{\mathrm{v}}^{\mathrm{v}} \end{bmatrix}\). The relation between these two forms is made explicit by writing \(\mathbf{F} = \mathrm{Id} + \delta \mathbf{F}\) which yields the component-wise identifications:
\begin{align*}
    \mathbf{F}_{\mathrm{h}}^{\mathrm{h}} &= \mathrm{Id} + \nabla u, &
    \mathbf{F}_{\mathrm{v}}^{\mathrm{v}} &= \mathrm{Id} + P, &
    \mathbf{F}_{\mathrm{h}}^{\mathrm{v}} &= N.
\end{align*}

In this coordinate setting, the deformation measures introduced in~\cite[p.~35, p.~39]{CrespoEtAlTwoScaleGeometricModelling2024} linearize as follows:
\begin{align*}
    \mathfrak{G}_{\mathrm{v}}^{\mathrm{v}} &= (\mathrm{Id} + P)^\mathrm{T} \cdot (\mathrm{Id} + P) &
    \boldsymbol\Theta &= (\mathrm{Id} + P)^{-1} \cdot (\mathrm{Id} + \nabla u) \\
    &\simeq \mathrm{Id} + 2 \op{sym}{P}, &
    &\simeq \nabla u - P, \\
    \boldsymbol\Gamma &= -(\mathrm{Id} + P)^{-1} \cdot N  &
    \boldsymbol\Gamma - \boldsymbol\Gamma_{\mathrm{holo}} &= -(\mathrm{Id} + P)^{-1} \cdot N + (\mathrm{Id} + P)^{-1} \cdot \nabla P \\
    &\simeq -N, &
    &\simeq \nabla P - N.
\end{align*}

\subsection{The Euler-Bernoulli beam}

The Euler-Bernoulli beam can be, in the bending case, seen as a bi-dimensional medium of classical elasticity under the two following assumptions: the beam hypothesis, namely that the transversal dimension is negligible compared to the longitudinal one; and the Euler-Bernoulli hypothesis:
\begin{align}
    \label{eq:EB_hypo}
    u_{\mathrm{E-B}}\left( X^1, X^2 \right) &:= \begin{bmatrix} -w'\left( X^1 \right) X^2 \\ w\left( X^1 \right)\end{bmatrix}.
\end{align}
To see this model as an instance of the micromorphic model one can, using the smallness of the transversal dimension \(X^2\), identify it with the microscopic scale \(Y\). That is, \({u_{\mathrm{E-B}}}^i_{,2} =: P^i_2\). Consequently, the longitudinal dimension \(X^1\) corresponds to the macroscopic dimension \(\overline X\). This, however, only provides the second column of \(P\) while leaving the first unspecified. In order to fully characterise \(P\) from \cref{eq:EB_hypo}, one can impose it to be antisymmetric. This last hypothesis can be seen as a manifestation of the transversal rigidity of the beam, an hypothesis used to derive the Euler-Bernoulli hypothesis. Doing so, one obtains:
\begin{align}
    u_{\mathrm{E-B}}\left( \overline X, Y \right)
        &:= \overline u_{\mathrm{E-B}}\left( \overline X \right) + P_{\mathrm{E-B}}\left( \overline X \right)\cdot \begin{bmatrix} 0 \\ Y \end{bmatrix}\\\nonumber
        &= \begin{bmatrix} 0 \\ w\left( \overline X \right)\end{bmatrix}
            + \begin{bmatrix} 0 & -w'\left( \overline X \right) \\ w'\left( \overline X \right) & 0 \end{bmatrix} \cdot \begin{bmatrix} 0 \\ Y \end{bmatrix}.
\end{align}

Notice that this formalism is almost holonomic, in the sense that the first column of \(P_{\mathrm{E-B}}\) is the longitudinal gradient of \(u\). The internal energy density of the Euler-Bernoulli beam, seen as an instance of classical elasticity, is a function of \(\opp{sym}{\nabla u_{\mathrm{E-B}}}\), where:
\begin{align}
    \nabla u_{\mathrm{E-B}}\left( \overline X, Y \right)
        &= \begin{bmatrix}
            -w''\left( \overline X \right) Y & \quad -w'\left( \overline X \right)\\
            w'\left( \overline X \right) & 0
        \end{bmatrix}.
\end{align}
That is, it is a function of the second derivative \(w''\):
\begin{align}
    \Psi_{\mathrm{E-B}}\left( \overline X, Y \right) &= \widehat\Psi\left( -w''\left( \overline X \right) Y \right).
\end{align}
Notice that this energy is also objective in the sense of micromorphic media~\cite{CrespoEtAlTwoScaleGeometricModelling2024}. As we are in the linear setting, \(\widehat{\Psi}\) is a quadratic form. Upon partial integration of the energy density along the transversal direction, one therefore obtains the following homogenised density:
\begin{align}
    \overline{\Psi}_{\mathrm{E-B}}\left( \overline X \right) &= E\,I_2 \left| w''\left( \overline X \right) \right|^2,
\end{align}
where \(I_2\) is the quadratic moment, transversal integral of \(Y^2\), and \(E\) is a constitutive scalar, called Young's modulus. In order to view the Euler-Bernoulli beam as an instance of our second-order model, one must specify \(N \in \mathbb{R}^{2\times2\times2}\). As the first column \(\at{N}_1\) is an avatar of \(\frac{\partial P}{\partial \overline X}\), one naturally sets \(\at{N}_1 = \frac{\partial P}{\partial \overline X}\). Using once again the smallness of the transversal dimension, one can neglect the second transversal derivatives. Hence, only \(N^1_{12}\), avatar of \(P^1_{1,2}=\frac{\partial P^1_1}{\partial Y}\), remains. Similarly to the micromorphic formalism, we choose the "most holonomic" value, which can be given by:
\begin{align}
    N^1_{12}
        \ \ 
            \simeq
        \ \ 
            P^1_{1,2}
        \ \ 
            \simeq
        \ \ 
             \left[ u_{\mathrm{E-B}} \right]^1_{,12}
        \ \ 
            \simeq
        \ \ 
            \left[ u_{\mathrm{E-B}} \right]^1_{,21}
        \ \ 
            \simeq
        \ \ 
             P^1_{2,1}
        \ \ 
            \simeq
        \ \ 
             -w''\left( \overline X \right),
\end{align}
leading to the following second-order Euler-Bernoulli hypothesis:
\begin{align}
    N\left( \overline X \right) &= \begin{bmatrix}
       \begin{bmatrix}
            0 & -w''\left( \overline X \right)\\
            w''\left( \overline X \right) & 0
       \end{bmatrix} 
       \begin{bmatrix}
        -w''\left( \overline X \right) & 0 \\ 0 & 0
       \end{bmatrix}
    \end{bmatrix}.
\end{align}
Notice that the model is still unidimensionally parametrised, in the sense that only one scalar field of one coordinate \(-\) namely \(w\left( \overline X \right)\) \(-\) is present. The internal energy density of the beam in this second-order formalism is, jumping directly to the homogenised version:
\begin{align}
\label{eq:EB_sec_order}
    \overline{\Psi}_{\mathrm{E-B}}\left( \overline X \right) &= E\,I_2 \left| w''\left( \overline X \right) \right|^2 + F\,I_4 \left| w'''\left( \overline X \right) \right|^2,
\end{align}
where \(I_4\) is the fourth-order moment, transversal integral of \(Y^4\), and \(F\) is a constitutive scalar. Once again, this energy is also objective in the sense of micromorphic media~\cite{CrespoEtAlTwoScaleGeometricModelling2024}. For brevity, we denote \(b := E\,I_2\) and \(c := F\, I_4\) the coefficients of the internal energy. We also denote \(l \in \mathbb{R}^*_+\) the (average) transversal thickness. In particular, one expects \(I_{2n}\) to be of the order of \(l^{2n+1}\).

An extension of the classical Euler-Bernoulli beam theory within the framework of strain gradient elasticity can be found in~\cite{LurieSolyaevRevisitingBendingTheories2018}. In this setting, one finds \(b \gg c\) and, more precisely, \(c\) is shown to scale like \(l^2\,b\). This is consistent with our model, if one assumes \(E\) and \(F\) to be of the same order. This approach builds on Mindlin's interpretation of microstructures~\cite{MindlinMicrostructureLinearElasticity1964}, but as the authors note, similar results can be derived using a simple gradient elasticity theory with surface tension~\cite[eq. 21 p. 389]{Papargyri-BeskouEtAlBendingStabilityAnalysis2003}. In the latter case, the governing equations converge to the same form in the limit where the surface term vanishes, \ie{} \(\ell \to 0\). In both cases, the resulting static equilibrium equations are:
\begin{align}
    \label{eq_euler_bern}
    f_0 &= b\,w^{(4)} - c\,w^{(6)} &&&&& \text{on \(\mathbb{B} = [0, L]\)},\\\nonumber
    T_{0} &= - b\,w^{(3)} + c\,w^{(5)} &&&&& \text{on \(\partial\mathbb{B} = \left\{ 0, L \right\}\)},\\\nonumber
    T_{1} &= b\,w^{(2)} - c\,w^{(4)} &
    &\text{and}&
    T_{2} &= c\,w^{(3)} & \text{on \(\partial\mathbb{B} = \left\{ 0, L \right\}\)},
\end{align}
which are exactly the ones obtained from \cref{eq:EB_sec_order} using variational calculus with the following total energy, assuming every coefficient is constant:
\begin{align}
    \overline{\boldsymbol\Psi} &= \int_0^L \overline{\Psi}_{\mathrm{E-B}} \left( \overline X \right) - f_0\, w\left( \overline X \right)\, \mathrm{d} \overline X
    + T_0(L)\,w(L) + T_1(L)\,w'(L)\\\nonumber
    &\qquad
     + T_2(L)\,w''(L) - T_0(0)\,w(0) - T_1(0)\,w'(0) - T_2(0)\,w''(0).
\end{align}

\subsection{Timoshenko beams}

The Timoshenko beam can be seen as a relaxation of the Euler-Bernoulli beam where \(P \in \mathfrak{so}(2)\) is still antisymmetric but independent of \(w\). That is, one has:
\begin{align}
    u_{\mathrm{Timo}}\left( \overline X, Y \right)
        &= u\left( \overline X \right) + P\left( \overline X \right) \cdot \begin{bmatrix} 0 \\ Y\end{bmatrix}\\\nonumber
        &= \begin{bmatrix}
            0 \\ w\left( \overline{X} \right)
        \end{bmatrix} + \begin{bmatrix}
            0 & -p\left( \overline{X} \right)\\
            p\left( \overline{X} \right) & 0
        \end{bmatrix} \cdot \begin{bmatrix}
            0 \\ Y
        \end{bmatrix}.
\end{align}

This time, the symmetric part of the gradient is less trivial:
\begin{align}
    \opp{sym}{\nabla u_{\mathrm{Timo}}}\left( \overline X, Y \right)
        &= \begin{bmatrix}
            -p'\left( \overline X \right) Y & w'\left( \overline X \right) - p\left( \overline X \right) \\
            w'\left( \overline X \right) - p\left( \overline X \right) & 0
        \end{bmatrix}.
\end{align}
Once again, the components are all objective quantities in the sense of \cite{CrespoEtAlTwoScaleGeometricModelling2024}. The Timoshenko beam model can be seen as an instance of our model by prescribing, using the same argument as earlier:
\begin{align}
    N &:= \begin{bmatrix}
        \begin{bmatrix}
             0 & -p'\\
             p' & 0
        \end{bmatrix}
        \begin{bmatrix}
            -n & 0\\
            0 & 0
        \end{bmatrix}
    \end{bmatrix}.
\end{align}
However, this time, \(N\) is an avatar of \(\nabla P\) but not of \(\nabla^2 u\), preventing the use of Schwartz equality which allowed us to set \(n = -p = w'\). One therefore has three independent scalar quantities. Once again, those are all objective quantities in the sense of \cite{CrespoEtAlTwoScaleGeometricModelling2024}. However, this is a peculiarity of this case, comming from the fact that most components are zero. Indeed, \cite{CrespoEtAlTwoScaleGeometricModelling2024} gives us in general \(n\), \(p'-n\) and \(u'-p\). We therefore choose to make the internal energy density of the beam quadratic in those. In the homogenised version, this yields:
\begin{align}
\label{eq:Timo_sec_order}
    \overline{\Psi}_{\mathrm{Timo}}\left( \,\cdot\,, Y \right)
        &= E\,I_2 \left| n \right|^2 + F\,I_4 \left| n' \right|^2
        + G\,I_0 \left| u' - p \right|^2 + H\, I_0 \left| p'-n \right|^2,
\end{align}

where \(G\) is the shear modulus, \(H\) is a constitutive scalar and \(I_0 \simeq l\) is the transversal width. This result can also be justified via an integration along \(X^2\). We naturally set \(d=G\, I_0\) and \(e = H\, I_0\). Classical Timoshenko beams have only the \(b:=E\, I_2\) and \(d\) terms. A generalisation that includes the case \( c := F\, I_4 \neq 0 \) is presented in~\cite[eq.~33]{WangEtAlMicroScaleTimoshenko2010}. While their model is broader in scope, it reduces to the semi-holonomic instance of the present framework \(-\) \ie{}, under the constraint \( n = p' \) \(-\) when the parameters are specified as \( \left( k_1, k_2, k_3, k_4, k_5 \right) = \left(c, b, 0, 0, d\right) \). This corresponds to the situation where they have \( l_1 = l_2 = 0 \), meaning there is no contribution from the deviatoric stretch gradient or the symmetric part of the rotation gradient. A comparable structure can also be derived from the analysis of a Timoshenko beam resting on a Pasternak foundation (\ie{} laying on a set of springs with shear interactions)~\cite{HarizEtAlBucklingTimoshenkoBeam2022}. The present authors have no knowledge of the existence of a model with \(c \neq 0\) and \(n \neq p'\) in the literature. Assuming that every coefficient is constant, variational calculus on \cref{eq:Timo_sec_order} yields:
\begin{align}
    f_0 &= d\left( u'' - p' \right) &&&&&&& \text{on \(\mathbb{B} = [0, L]\)},\\\nonumber
    f_1 &= d\left( u' - p \right) - e \left( p'' - n' \right) \hspace{-5em}&&&&&&& \text{on \(\mathbb{B} = [0, L]\)},\\\nonumber
    f_2 &= b\, n - c\, n'' &&&&&&& \text{on \(\mathbb{B} = [0, L]\)},\\\nonumber
    T_0 &= d\left( u'-p \right), &
    T_1 &= e\left( p'-n \right) &
    &\text{and}&
    T_2 &= c\, n' & \text{on \(\partial\mathbb{B} = \left\{ 0, L \right\}\)},
\end{align}
where \(f_0/T_0\), \(f_1/T_1\) and \(f_2/T_2\) are bulk/boundary forces acting on \(u\), \(p\) and \(n\) respectively.

\section{Defects and behaviour of the second-order Euler-Bernoulli beam}

In the linearized theory of microstructured continua, torsion and curvature tensors emerge as fundamental descriptors of defects. The \emph{torsion tensor} represents the density of dislocations---line defects where the material's internal structure fails to close after a loop---and is algebraically captured by the skew-symmetric part of the connection \( \boldsymbol \Gamma \). More precisely, in the linear regime, the torsion tensor \( T^i_{jk} \) is given by the antisymmetric combination \( T^i_{jk} = N^i_{jk} - N^i_{kj} \). On the other hand, the \emph{curvature tensor} encodes the density of disclinations, rotational defects resulting from incompatibilities in the material's micro-rotation field. In the linearised form, it is expressed as the curl of \( N \), namely \( R^i_{jkl} = N^i_{jk,l} - N^i_{jl,k} \).

Both tensors are antisymmetric in their last indices. In two dimensions, this means that they are entirely determined by \(T^i_{12}\) and \(R^i_{j12}\) respectively. In the case of the proposed second-order generalisation of the Euler-Bernoulli beam, one has:
\begin{align}
    T^i_{12}
        &= N^i_{12} - N^i_{21}
        = \delta^i_1\left( -n + p' \right),
\end{align}

meaning that, in this formalism, no dislocation can arise in an Euler-Bernoulli beam. On the other-hand, the curvature field follows:
\begin{align}
    R^i_{j12}
        &= N^i_{j1,2} - N^i_{j2,1}
        = - N^i_{j2,1}
        = \delta^i_1\delta^1_j\,n'.
\end{align}

That is, for an Euler-Bernoulli beam, the disclinations density field is proportional to the third order derivative \(u^2_{,111} = w'''\) of the transversal displacement.

\subsubsection*{Induced defect equations \(-\) Euler-Bernoulli}

From the ordinary differential system \cref{eq_euler_bern}, it follows that the curvature \(R^1_{112}\) of an Euler-Bernoulli beam follows:
\begin{align}
    f_0 &= b\,{R^1_{112}}^{(1)} - c\,{R^1_{112}}^{(3)} &&&&& \text{on \(\mathbb{B} = [0, L]\)},\\\nonumber
    T_{0} &= - b\,{R^1_{112}} + c\,{R^1_{112}}^{(2)}&
    &\text{and}&
    T_{2} &= c\,{R^1_{112}} & \text{on \(\partial\mathbb{B} = \left\{ 0, L \right\}\)}.
\end{align}

This is a Helmholtz equation in \({R^1_{112}}'\), meaning \({R^1_{112}}' = f_0 \star G\) with \(G(r) = \frac {\sqrt c}{2\sqrt b}\, e^{-\sqrt{\frac{b}{c}}|r|}\). From Young's inequality one therefore has
\begin{align}
    \frac{\left\| {R^1_{112}}^{(1)} - f_0 \right\|_\infty}{\left\| f_0 \right\|_\infty} &\leq \left\| G - \delta \right\|_1 = \frac cb \ll 1.
\end{align}
This means that \({R^1_{112}}^{(1)}\) is relatively close to \(f_0\).

\subsubsection*{Induced defect equations \(-\) Timoshenko}

For a Timoshenko beam, let us denote the relative orientation of the fibre by \(\theta := u' - p\). The system of equations of a Timoshenko beam then becomes:
\begin{align}
    f_0 &= d\, \theta^{(1)} & \text{on \(\mathbb{B} = [0, L]\)},\\\nonumber
    T_0 &= d\, \theta & \text{on \(\partial\mathbb{B} = \left\{ 0, L \right\}\)},\\
    f_1 &= d\, \theta - e\,{T^1_{12}}^{(1)} & \text{on \(\mathbb{B} = [0, L]\)},\\\nonumber
    T_1 &= e\, T^1_{12} & \text{on \(\partial\mathbb{B} = \left\{ 0, L \right\}\)},\\
    f_2' &= b\, {R^1_{112}} - c\, {R^1_{112}}^{(2)} & \text{on \(\mathbb{B} = [0, L]\)},\\\nonumber
    T_2 &= c\, {R^1_{112}} & \text{on \(\partial\mathbb{B} = \left\{ 0, L \right\}\)},
\end{align}

where each group can be seen as an independent system if \(\theta\) is fixed in the second group; which can be done by solving it using the first two equations. Importantly, this time one has a Helmholtz equation in \({R^1_{112}}\) where, using the same arguments as above, \({R^1_{112}}\) is relatively close to \(f_2'\).

\subsubsection*{Conclusion} In this article, it has been demonstrated how the classical models of Euler-Bernoulli and Timoshenko beams can be generalised to a second-order non-holonomic formalism. Doing so, it was shown that the generalised Euler-Bernoulli model exhibit no torsion, but a curvature linked to a third order derivative of the displacement. Parallelly, the generalised Timoshenko beam was shown to exhibit both curvature and torsion, where the former corresponds to the non-holonomy introduced in the generalisation. Lastly, using variational calculus, asymptotic values when \(c \ll b\) for the curvature in static equilibrium where obtained, as long as an equation for the torsion in the generalised Timoshenko beam.

\begin{credits}
\subsubsection{\ackname}
Mewen Crespo would like to thank the Faculty of Mathematics of the University of Duisburg-Essen for its hospitality during the semester-long on-site collaboration with Patrizio Neff, of which the present work is the fruit. Furthermore, the authors would like to thank the CNRS' research group GDR GDM CNRS for stimulating the interactions between differential geometry and mechanics. An interaction which is at the heart of this work. They would also like to thank CNRS, Univ Rennes, ANR-11-LABX-0020-0 program Henri Lebesgue Center for their financial support.

\subsubsection{\discintname}
The authors have no competing interests to declare that are relevant to the content of this article.
\end{credits}
%
%
\bibliographystyle{splncs04}

\begin{thebibliography}{8}

\bibitem{VolterraTheoryDislocations1907}
Volterra, V.: Sur l'équilibre des corps élastiques multiplement connexes. \textit{Annales Scientifiques de l'École Normale Supérieure} \textbf{24}, 401--517 (1907)   

\bibitem{MindlinMicrostructureLinearElasticity1964}
Mindlin, R.D.: Micro-structure in linear elasticity. \textit{Archive for Rational Mechanics and Analysis} \textbf{16}, 51--78 (1964). \doi{10.1007/BF00248490}

\bibitem{EringenSuhubiNonlinearTheorySimple1964}
Eringen, A.C., Suhubi, E.S.: Nonlinear theory of simple micro-elastic solids I. \textit{International Journal of Engineering Science} \textbf{2}(2), 189--203 (1964). \doi{10.1016/0020-7225(64)90004-7}

\bibitem{ToupinElasticMaterialsCouplestresses1962}
Toupin, R. A.: Elastic materials with couple-stresses. \textit{Archive for Rational Mechanics and Analysis} \textbf{11}(1), 385--414 (1962)

\bibitem{NeffEtAlUnifyingPerspectiveRelaxed2014}
Neff, P., Ghiba, I.-D., Madeo, A., Placidi, L., Rosi, G.: A unifying perspective: the relaxed linear micromorphic continuum. \textit{Continuum Mechanics and Thermodynamics} \textbf{26}(5), 639--681 (2014)

\bibitem{NeffForestGeometricallyExactMicromorphic2007}
Neff, P., Forest. S.: A geometrically exact micromorphic model for elastic metallic foams accounting for affine microstructure. \textit{Journal of Elasticity} \textbf{87}(2-3), 239--276 (2007)

\bibitem{CrespoEtAlTwoScaleGeometricModelling2024}
Crespo, M., Casale, G., Le Marrec, L.: Two-scale geometric modelling for defective media. \textit{Mathematics and Mechanics of Solids} (2025), in press (preprint: \url{https://arxiv.org/abs/2404.03269})

\bibitem{CrespoEtAlContinuumMechanicsDefective2023}
Crespo, M., Casale, G., Le Marrec, L.: Continuum mechanics of defective media: an approach using fiber bundles. In: Nielsen, F., Barbaresco, F. (eds.) \textit{Geometric Science of Information}, LNCS, vol. 14072, pp. 41--49. Springer, Cham (2023). \doi{10.1007/978-3-031-38299-4_5}

\bibitem{CartanGeneralisationNotionCourbure1922}
Cartan, É.: Sur une généralisation de la notion de courbure de Riemann et les espaces à torsion. \textit{Comptes Rendus de l'Académie des Sciences de Paris} \textbf{174}, 593--595 (1922)

\bibitem{LurieSolyaevRevisitingBendingTheories2018}
Lurie, S., Solyaev, Y.: Revisiting bending theories for gradient elastic beams. \textit{International Journal of Engineering Science} \textbf{126}, 1--21 (2018). \doi{10.1016/j.ijengsci.2018.01.002}  

\bibitem{Papargyri-BeskouEtAlBendingStabilityAnalysis2003}
Papargyri-Beskou, S., Beskos, D.E., Polyzos, D.: Bending and stability analysis of gradient elastic beams. \textit{International Journal of Solids and Structures} \textbf{40}(2), 385--400 (2003). \doi{10.1016/S0020-7683(02)00522-X}

\bibitem{WangEtAlMicroScaleTimoshenko2010}
Wang, B., Zhao, J., Zhou, S.: A micro scale Timoshenko beam model based on strain gradient elasticity theory. \textit{European Journal of Mechanics} \textbf{29}(4), 591--599 (2010). \doi{10.1016/j.euromechsol.2009.12.005}

\bibitem{HarizEtAlBucklingTimoshenkoBeam2022}
Hariz, M., Le Marrec, L., Lerbet, J.: Buckling of Timoshenko beam under two-parameter elastic foundations. \textit{International Journal of Solids and Structures} \textbf{244-245} (2022). \doi{10.1016/j.ijsolstr.2022.111583}
\end{thebibliography}
%
\renewcommand{\doi}[1]{}

\end{document}